\documentclass[aps,prd,12pt,a4paper,groupedaddress,preprintnumbers,floatfix,nofootinbib,showpacs]{revtex4-1}
\usepackage[english]{babel}
\usepackage{amsmath,amssymb}
\usepackage{graphicx}
\usepackage{color}
\usepackage{bbold,mathbbol}
\usepackage{hyperref}
\usepackage{dcolumn}
\usepackage{bm}
\begin{document}
\preprint{DCP-12-03}

\title{A $Z_4$ flavor model in RS1}

\author{
  Carlos Alvarado,$^{1}$\footnote{Electronic address:  arcarlos00@gmail.com}
Alfredo Aranda,$^{1,2}$\footnote{Electronic address: fefo@ucol.mx}
Olindo Corradini,$^{3}$\footnote{Electronic address: olindo.corradini@gmail.com} 
Alma D. Rojas,$^{1}$\footnote{Electronic address: alma.drp@gmail.com}
Eli Santos-Rodriguez$^{4}$\footnote{Electronic address: eli@unach.mx}  \vspace*{0.3cm}}

\affiliation{$^1$Facultad de Ciencias, CUICBAS,
Universidad de Colima, Colima, M\'exico \\
$^2$Dual C-P Institute of High Energy Physics, M\'exico \\
$^3$ Centro de Estudios en F\'isica y Matem\'aticas B\'asicas y Aplicadas\\  Universidad Aut\'onoma de Chiapas, Tuxtla Guti\'errez, Chiapas, M\'exico \\
$^4$ International Centre for Theoretical Physics,
Meso-American Institute for Science\\ Universidad Aut\'onoma de Chiapas, Tuxtla Guti\'errez, Chiapas, M\'exico}

\date{\today}


\begin{abstract}
Randall Sundrum models provide a possible explanation of (gauge-gravity) hierarchy, whereas discrete symmetry flavor groups yield a possible description of the texture of Standard Model fermion masses. We use both these ingredients to propose a five-dimensional extension of the Standard Model where the mass hierarchy of the four-dimensional effective field theory is obtained only using localizations parameters of order 1. We consider a bulk custodial gauge symmetry group together with an Abelian $Z_4$ group: the model turns out to yield a rather minimal extension of the SM as it only requires two brane Higgs fields to provide the desired Yukawa interactions and the required spontaneous symmetry breaking pattern.  In fact, the presence of an extra-dimension allows the use of the Scherk-Schwarz mechanism to contribute to the breaking of the bulk custodial group down to the SM gauge symmetry. Moreover, no right-handed neutrinos are present and neutrino masses are generated radiatively with the help of a bulk charged scalar field that provides the Lepton-number violation. Using experimental inputs from the Global Neutrino Analysis and recent Daya Bay results, a numerical analysis is performed and allowed parameter regions are displayed.
\end{abstract}


\maketitle

\section{Introduction}
\label{sec:intro} 

The Standard Model (SM) gauge group is {\it a priori} consistent with a large flavor group that is not observed experimentally and Yukawa couplings and mixings are introduced to comply with such experimental evidence~\cite{Cabibbo:1963yz}. However, the pattern of fermion masses and mixing angles and the nature of the neutrino mass (whether Dirac or Majorana) is not constrained by the gauge symmetry and it is thus natural to go beyond the SM to seek for a theoretical explanation of the observed masses and angles and of their hierarchies. To date a promising direction towards the understanding of flavor physics in the SM -- notably of fermion mass matrices and mixing -- has been the study of particular textures (in the quark sector), such as the Nearest Neighbor Interactions (NNI) texture~\cite{Branco:2010tx,Harayama:1996am,NNI} and the Fritzsch type textures~\cite{Fritzsch:1977vd}, and the introduction of flavor symmetry groups that constrain the structure of the fermion mass matrix to reproduce the desired texture. The mixing in the lepton sector has also given strong support to the idea of an underlying flavor symmetry and many flavor symmetry groups have been proposed so far~\cite{Altarelli:2010gt,Ishimori:2010au,Altarelli:2010fk}; of them discrete groups have proved to be quite successful and we keep exploring such direction in the present work. 

Although the aforementioned approach, involving (discrete) flavor groups, provides a possible description of the pattern of fermion masses and mixing angles, it seems not to be able to explain the hierarchy(ies) among the different species of fermionic masses. i.e., the flavor symmetry explains the number and location of the "zeros" in the textures for the fermion mass matrices, but not the relative size among the non-zero entries. In other words it appears that, in order to obtain experimentally suitable values for such physical quantities, one needs to allow for parameters of quite different order of magnitude. Such hierarchy problems have proved to be quite hard, in the framework of four-dimensional non-supersymmetric theories. On the other hand in the context of extra-dimensions, several models -- with flat or warped extra-dimensions -- were proposed as solutions to hierarchy problems, mostly focusing on the gauge-gravity hierarchy.  In particular the Randall-Sundrum model 1 (RS1)~\cite{Randall:1999ee}  has drawn a great deal of attention as it provides a possible description of the TeV-Planck hierarchy with a small (Planck size) extra-dimension and treats the TeV scale as a derived scale. The geometry of such space is a slice of five-dimensional Anti-de Sitter space, a warped product between four-dimensional space, and an extra dimension with the topology of an interval whose extrema are the fixed points of a $Z_2$ orbifold symmetry. The space thus "ends" on two flat four-dimensional planes (branes) located at the extrema of the extra dimension. Extra-dimensional models, especially Randall-Sundrum models, have also been explored in the context of the SM physics as fermion masses might be explained in terms of the overlaps between the extra-dimensional profiles of fermions and Higgs(es)~\cite{Grossman:1999ra,ArkaniHamed:1999dc,Gherghetta:2000qt,Huber:2000ie,Huber:2003tu}. 

It is then interesting to consider the possibility of merging both scenarios -- flavor symmetries and hierarchy from extra dimensional settings -- in a single setup and to determine if a viable model can be constructed. This work presents one such case. To do this we use a minimally-extended version of the SM that incorporates a discrete flavor group, and that reproduces fermion mass patterns as well as mixing angles in both the quark and lepton sectors. The model is minimal in the sense that it contains the SM matter fields only (in particular, it does not add right-handed neutrinos), an extended Higgs sector composed of two Higgs SU(2) doublets, a Lepton number violating scalar, and the smallest Abelian group that renders the NNI textures for the quark mass matrices, namely the cyclic group $Z_4$. 

The model is detailed in the forthcoming section, however let us stress here that such approach was already taken by other groups. For example, in~\cite{Chen:2008qg}, a realization of the so-called Lepton Minimal Flavor Violation scenario was obtained in the framework of RS with right-handed neutrinos. There it was found that the first Kaluza-Klein (KK) mass scale could be as low as $3$~TeV. Along similar lines, but for the quark sector and providing a solution to the issue of flavor changing neutral currents in extra dimensional settings, the work in~\cite{Bauer:2011ah} extends the strong sector in the bulk with an additional SU(3), broken down to Quantum Chromodynamics (QCD) by boundary conditions that render the model safe, even for a KK scale of $2$~TeV. Closer to our approach is the work by Kadosh and Pallante~\cite{Kadosh:2010rm}, where an $A_4$ flavor symmetry is introduced in a RS setup. The main difference with our model is that, unlike~\cite{Kadosh:2010rm}, we use a flavor group that reproduces the mixing matrices in both the quark and lepton sectors automatically, without the need of introducing additional flavon fields to generate non-zero entries and/or further suppressions. This makes our model more economical in terms of additional fields and more tractable, specially in the scalar sector. Another important difference is that in our model neutrino masses are generated radiatively and there is no need to introduce right-handed neutrinos, as mentioned above.

The gauge and Yukawa sectors of our model are presented in section~\ref{sec:model}, where the expressions for the mass matrices in both the lepton and quark sectors are described and explicitly shown. Section~\ref{sec:numerical} shows the numerical analysis and results for both sectors and some remarks about the scalar potential and flavor changing neutral currents. Finally we present the conclusions. An appendix has been included in order to maintain the discussion in the paper self contained.

\section{Model}
\label{sec:model} 

The five-dimensional field theory we consider is a higher-dimensional extension of the SM and lives in the RS1 background. Such space is a slice of AdS$_5$ space of Planck-size curvature, where the fifth dimension $y$ is taken to be an orbifolded circle $S_1/Z_2$ of radius $R$,  and fields are odd or even under the action of the orbifold, i.e. under reflection $y\to -y$. In other words the extra dimension has the topology of a line that stretches between $y=0$ and $y=\pi R$ and the latter are fixed points of the orbifold, where two flat 3-branes, one with positive and the other with negative tension, are accommodated. The brane located at $y=0$ is referred to as the "UV brane", whereas the brane located at $y=\pi R$ is referred to as the "IR brane". The bulk line element is thus given by
\begin{equation}
ds^2 = dy^2 +e^{-2k|y|} \eta_{\mu\nu} dx^\mu dx^\nu\,,\qquad -\pi R\leq y\leq \pi R 
\label{eq:RS1}
\end{equation}   
 with $k\sim M_{Planck}$ being (proportional to) the AdS curvature, and the TeV-Planck hierarchy is realized provided $k R\approx 10$. Such geometry is realized with a negative bulk cosmological constant and with the aforementioned brane tensions: all three are of Planck size. Moreover, all the fields we deal with have vacuum expectation values (vevs) of order at most TeV, and we may thus neglect their backreaction to the geometry~\eqref{eq:RS1}.  
 
The five-dimensional model we consider has a gauge-symmetry group that includes the Standard Model custodial symmetry $G_{CS} = SU(3)_C \times SU(2)_L \times SU(2)_R \times U(1)_{B-L}$ in order to suppress excessive contributions to the Peskin-Takeuchi $T$ parameter~\cite{Peskin:1990zt,Agashe:2003zs}, augmented with a discrete Abelian group $Z_4$, hence $G=G_{CS} \times Z_4$.  All the fields of our model are charged under $Z_4$ and, with the exception of the Higgs fields, are all bulk fields, along the lines of what is done in~\cite{Agashe:2003zs}. This setup thus enjoys a very nice interpretation in terms of AdS/CFT correspondence as only the Higgs fields are seen as TeV-scale composites of the strongly-coupled sector of the four-dimensional theory and, since all other fields are bulk fields, all the phenomenology of the model is addressable from the weakly-coupled five-dimensional model. 

The matter content of the model involves (three families of) left-handed quark doublets $Q_{Li}$ and lepton doublets $L_{Li}$ along with two copies of right-handed quark doublets, $U_{Ri} =\scriptstyle{\left(\begin{array}{l} u_{Ri}\\ \tilde d_i \end{array}\right) }$ and $D_{Ri} =\scriptstyle{\left(\begin{array}{l} \tilde u_{i}\\ d_{Ri}\end{array}\right)}$, and one copy of right-handed lepton doublets, $E_{Ri} =\scriptstyle{\left(\begin{array}{l} \tilde\nu_{i}\\ l_{Ri}\end{array}\right)}$; the zero-modes of "tilded" fields will be projected out from the IR brane by Scherk-Schwarz gauge-symmetry breaking~\cite{Agashe:2003zs} and only their KK modes are eventually non-zero on the brane.  We then use two Higgs fields $\Phi_1$ and $\Phi_2$ that are bi-doublets of $SU(2)_L\times  SU(2)_R$ and are confined to the IR brane. Lastly, we introduce a bulk singlet scalar field $h$ charged under hypercharge and lepton number that participates in neutrino mass generation. The $Z_4$ charge assignment for all such fields is similar to the one given in~\cite{Aranda:2011rt}, namely
\begin{eqnarray}
(q_1,q_2,q_3) &=& (2,0,3)\\
(u_1,u_2,u_3) &=& (d_1,d_2,d_3) = (3,1,2)
\end{eqnarray}   
for the quark sector, with $q_i = Q(Q_{Li})$, $u_i=Q(U_{Ri})$ and  $d_i=Q(D_{Ri})$,
\begin{eqnarray}
(\alpha_1,\alpha_2,\alpha_3) &=& (2,0,3)\\
(e_1,e_2,e_3) &=& (3,1,2)
\end{eqnarray}   
for the lepton sector,  with $\alpha_i = Q(L_{Li})$ and  $e_i=Q(E_{Ri})$, and
\begin{eqnarray}
(\phi_1,\phi_2) = (1,2) \ , \ \ \ (\chi) = 1
\end{eqnarray}  
for the scalars with $\phi_i = Q(\Phi_i)$, and $\chi = Q(h)=1$. The main difference with~\cite{Aranda:2011rt} is that the left-handed up-type and down-type quarks have the same $Z_4$ charges; the reason is dictated by the fact that we are using Higgs bi-doublets confined on the IR brane. These fields transform as $\Phi' = e^{i\lambda_{aL} \tau_{{aL}}}\ \Phi\ e^{-i\lambda_{aR} \tau_{{aR}}}$ under $SU(2)_L\times SU(2)_R$ gauge transformations ($\tau_{aR} = \tau_{aL} = \frac{1}{2} \sigma_a$, where $\sigma_a$ are the Pauli matrices) and, as explained later, are responsible for the spontaneous breaking of $SU(2)_L\times SU(2)_R \to SU(2)_D$ on the IR brane. In other words the vevs of $\Phi_i$ are $SU(2)_D$ singlets
\begin{equation}
\big< \Phi \big> =\frac{v}{\sqrt{2}}\left[\left(\begin{array}{c} 1\\ 0\end{array}\right)\otimes \left(\begin{array}{c c} 1 & 0\end{array}\right) + \left(\begin{array}{c} 0\\ 1\end{array}\right)\otimes \left(\begin{array}{c c} 0 & 1\end{array}\right)\right]
\label{eq:vev}
\end{equation}
 so that $\big<\tilde \Phi_i\big> = \big<\Phi_i$\big>, with $\tilde \Phi \equiv (i\sigma_{_{2}}) \Phi^* (-i\sigma_{_{2}})$, and $U_R$ and $D_R$ thus enter on equal footing in the Yukawa lagrangian and must then have the same $Z_4$ charges. 
 
 It is easy to check that~\eqref{eq:vev} is invariant under the diagonal part of  $SU(2)_L\times SU(2)_R$, namely
 \begin{equation}
 \delta \big< \Phi \big> = i \lambda_a \Big ( \tau_{_{aL}} \big< \Phi \big> -\big< \Phi \big> \tau_{_{aR}} \Big) =0~.
\end{equation}  
and thus preserves $SU(2)_D$.
 
\subsection{Gauge-symmetry breaking} 
\label{sec:gauge-symmetry-breaking}
The bulk (custodial) gauge symmetry must be broken down to the SM gauge group on the four-dimensional effective action at TeV scale and usual electroweak spontaneous symmetry breaking must also be accounted for. We realize this in the same way as in~\cite{Agashe:2003zs,Kadosh:2010rm} by orbifold Scherk-Schwarz projection, i.e. by assigning independent orbifold charges to a field at the two different ends of the interval, and with canonical spontaneous breaking. 

We can impose on the bulk  fields a $Z_2 \times Z_2' $ charge assignment. For scalar fields this corresponds to the field transformations  
\begin{eqnarray}
\phi(x,-y) &=& Z \phi(x,y)\\
\phi(x,\pi R -y) &=& Z' \phi(x,\pi R+y) ~,  
\end{eqnarray}
with $Z,Z'=\pm$ and $y = 0, \pi R$ being the fixed points of  $Z_2 \times Z_2' $. For 5D spinors, single valuedness of the lagrangian upon the action of $Z_2 \times Z_2'$ requires 
\begin{equation}
\label{eq:SS-fermions}
\begin{split}
& \psi(x,y) = Z \gamma_5 \psi(x,-y)\\
& \psi(x,\pi R+y) = Z' \gamma_5 \psi(x,\pi R-y) ~.
\end{split}
\end{equation}
  We can fix $\gamma_5 =\scriptstyle{\left( \begin{array}{cc} {\mathbb 1} & 0\\
0 & -{\mathbb 1} \end{array}\right)}$ and decompose the 5D 4-spinor as $\psi = \scriptstyle{\left( \begin{array}{c} \xi\\
\eta \end{array}\right)}$ with $\xi (\eta)$ being left-handed (right-handed) Weyl spinors, in the 4D sense. At the fixed points, i.e. setting $y=0$ in~\eqref{eq:SS-fermions}, we thus have that the left-handed Weyl spinor $\xi$ has charges $(Z,Z')$ and the corresponding right-handed Weyl spinor $\eta$ has charges $(-Z,-Z')$~\footnote{Henceforth, when referring to the fermionic $Z_2 \times Z_2' $ charges, we will indicate the charges associated to the left-handed part of the doublet.}.  Below we mostly only care about the extra-dimensional zero-modes, whose profiles are summarized in Appendix~\ref{appendix:profiles}.  We thus have that the fermion left-handed zero mode~\eqref{eq:zero-modeF} only exists for charges $(Z,Z')=(+,+)$ whereas the right-handed zero mode exists for $(Z,Z')=(-,-)$. In other words, other non-trivial orbifold projections lift the mass of the zero-mode. For vector fields, whose would-be zero-modes are constant, we again have that only  $(Z,Z')=(+,+)$ allows for massless zero modes and other charge assignments lift the mass and thus realize the low-energy symmetry breaking.  

In our model we break $SU(2)_R$ gauge symmetry down to $U(1)_R$ on the UV brane via Scherk-Schwarz mechanism, i.e. we assign $(Z,Z')=(-,+)$ to the electrically charged vector bosons of $SU(2)_R$, and $(Z,Z')=(+,+)$ to the neutral gauge boson. Since the former couple the upper and lower parts of the right-handed fermion doublets, single-valuedness of the bulk lagrangian requires that if the upper part is even, on the UV brane, the lower part must be odd and viceversa. Hence, in order to have zero-modes for both the upper and lower parts of the right-handed quarks, we need to double the number of right-handed fields~\cite{Agashe:2003zs} in the quark sector, as already mentioned above. The charge assignment for such fields thus reads
\begin{equation}
Q_{Li}[+,+]\,,\qquad U_{Ri} =\left(\begin{array}{l} u_{Ri}[-,-]\\ \tilde d_i[+,-] \end{array}\right) \,,\qquad D_{Ri} =\left(\begin{array}{l} \tilde u_{i}[+,-]\\ d_{Ri}[-,-]\end{array}\right)
\end{equation}
for the quark sector, and   
\begin{equation}
L_{Li}[+,+]\,,\qquad E_{Ri} =\left(\begin{array}{l} \tilde \nu_{i}[+,-]\\ e_{Ri}[-,-]\end{array}\right)
\end{equation}
for the leptonic sector. A vev on the UV brane then provides the breaking $U(1)_R\times U(1)_{B-L} \to U(1)_Y$~\cite{Agashe:2003zs}.

On the IR brane the vevs of the Higgs bi-doublets provide the spontaneous breaking $SU(2)_L\times SU(2)_R \times U(1)_{B-L}\to SU(2)_D\times U(1)_{B-L}$. Hence finally the superposition of all such breakings only leaves $U(1)_{\rm em}$ untouched. In fact
\begin{equation}
\underbrace{\tau_{_{3L}} +\tau_{_{3R}}}_{\tau_{_{3D}}} +\frac{1}{2}(B-L) = \tau_{_{3L}} + Y = Q_{em} 
\end{equation}  
with $\tau_{aD} = \tau_{_{aL}}\otimes {\mathbb 1}_R + {\mathbb 1}_L\otimes\tau_{_{aR}}$ being the generators of $SU(2)_D$ (above the tensor product is left implied.)

\subsection{Yukawa Sector}
\label{subsec:yukawa}

\subsubsection{Quarks}
\label{subsubsec:quark}

The mass terms for the quarks come from the following Yukawa interactions on the IR brane
\begin{eqnarray}
\label{eq:yukawa-quark} \nonumber
 -\mathcal{L}_{\text{Yukawa}}^{q} &=& \int \text{d}y\dfrac{\sqrt{-g}~\delta(y-\pi R)}{\Lambda}\biggl[ (\gamma_{u}^{1})_{ij}\bar{Q}_{Li}(x^{\mu},y)\Phi_{1}(x^{\mu})U_{Rj}(x^{\mu},y) \biggr.  \\
                                  &+& (\gamma_{u}^{2})_{ij}\bar{Q}_{Li}(x^{\mu},y)\Phi_{2}(x^{\mu})U_{Rj}(x^{\mu},y)+(\gamma_{d}^{1})_{ij}\bar{Q}_{Li}(x^{\mu},y)\Phi_{1}(x^{\mu})D_{Rj}(x^{\mu},y) \nonumber \\
                                  &+& \biggl. (\gamma_{d}^{2})_{ij}\bar{Q}_{Li}(x^{\mu},y)\Phi_{2}(x^{\mu})D_{Rj}(x^{\mu},y) \biggr] + h.c.,
\end{eqnarray}
where $\Lambda \equiv M_{Pl}$ is the Planck-scale and $(\gamma_{u,d}^{1,2})_{ij}$ are dimensionless parameters assumed of $\mathcal{O}(1)$ that, together with the quark extradimensional profiles, generate the effective four-dimensional Yukawa couplings. The $Z_4$ charge assignments for the quark and scalar fields then induce the desired Yukawa NNI textures (writing separately each scalar contribution)
\begin{eqnarray} \label{gammas}
 \gamma_{u,d}^{1} &=& \left(\begin{array}{ccc}
                0    & \ast & 0 \\
                \ast & 0    & 0 \\
                0    & 0    & \ast
                \end{array}\right),~~~~~
 \gamma_{u,d}^{2}=\left(\begin{array}{ccc}                                                                       0 & 0    & 0 \\                                                                       0 & 0    & \ast \\                                                                       0 & \ast & 0                                                                     \end{array}\right),
\end{eqnarray}
which after electroweak symmetry breaking lead to the following quark mass matrices
\begin{equation}
 \label{eq:Z4quarkmassmatrix}
 M_{u}=\left(\begin{array}{ccc}
              0                                      & \Gamma_{12}^{u}\langle \Phi_{1}\rangle & 0 \\
              \Gamma_{21}^{u}\langle \Phi_{1}\rangle & 0                                      & \Gamma_{23}^{u}\langle \Phi_{2}\rangle \\
              0                                      & \Gamma_{32}^{u}\langle \Phi_{2}\rangle & \Gamma_{33}^{u}\langle \Phi_{1}\rangle
             \end{array}\right),~~~~~ M_{d}=\left(\begin{array}{ccc}
                                                   0                                      & \Gamma_{12}^{d}\langle \Phi_{1}\rangle & 0 \\
                                                   \Gamma_{21}^{d}\langle \Phi_{1}\rangle & 0                                      & \Gamma_{23}^{d}\langle \Phi_{2}\rangle \\
                                                   0                                      & \Gamma_{32}^{d}\langle \Phi_{2}\rangle & \Gamma_{33}^{d}\langle \Phi_{1}\rangle
                                                  \end{array}\right)
\end{equation}
with $\langle \Phi_{i}\rangle$ denoting the vevs. The $\Gamma_{ij}^{u,d}$ are the effective four-dimensional Yukawa couplings that depend on the fermion extradimensional profiles overlap with the Higgs bi-doublets at the IR boundary. Thus, under the zero mode approximation (ZMA) (see Appendix for the KK decomposition and the explicit fermion profiles), each of the Yukawa terms above looks like
\begin{eqnarray} \nonumber
 -\mathcal{L}_{q} &\supset & \int_{-\pi R}^{\pi R}\text{d}y~\dfrac{e^{-4k|y|}\delta(y-\pi R)}{\Lambda}\gamma_{ij}\dfrac{\bar{Q}_{Li}^{(0)}(x)f_{QLi}^{(0)}(y)}{\sqrt{2\pi R}}\Phi(x) \dfrac{Q_{Rj}(x)f_{QRj}^{(0)}(y)}{\sqrt{2\pi R}}e^{\pi kR} \\
&=& \left\{\gamma_{ij}\dfrac{k}{\Lambda}\sqrt{\dfrac{(1/2-c_{Li}^{Q})(1/2-c_{Rj}^{q})}{[e^{2\pi kR(1/2-c_{Li}^{Q})}-1][e^{2\pi kR(1/2-c_{Rj}^{q})}-1]}}e^{(1-c_{Li}^{Q}-c_{Rj}^{q})\pi kR} \right\}\bar{Q}_{Li}^{(0)}(x)\Phi(x)Q_{Rj}^{(0)}(x), \nonumber \\
\end{eqnarray}
where $q=U,D$, $\Phi=\Phi_{1,2}$, and $i$, $j$ are family indices accordingly chosen (note also the inclusion of a canonically-normalizing factor for the Higgs bi-doublet). The effective Yukawa coupling, given by the product inside the curly brackets, depends on two quark localization $c$~-- parameters
\begin{equation}
\label{eq:4DYukawaquarks}
\Gamma_{ij}^{u,d}=(\gamma_{u,d})_{i,j}\dfrac{k}{\Lambda}\sqrt{\dfrac{(1/2-c_{Li}^{Q})(1/2-c_{Rj}^{u,d})}{[e^{2\pi kR(1/2-c_{Li}^{Q})}-1][e^{2\pi kR(1/2-c_{Rj}^{u,d})}-1]}}e^{\left(1-c_{Li}^{Q}-c_{Rj}^{u,d}\right)\pi kR}\ ,
\end{equation}
where $c_{Li}^{u}=c_{Li}^{d}\equiv c_{Li}^{Q}$ since the left handed components of the $u$ and $d$ quarks form an $SU(2)_{L}$ doublet. In order to extract the $c$~-- parameters that lead to experimentally allowed observables, we follow the Harayama parametrization in reference \cite{Harayama:1996am}, which is a transformation of the up -- and down -- type quark mass matrices to a basis such that these display the NNI form without modifying the mass eigenvalues nor the CKM matrix entries. Once parametrized in that form and following~\cite{Aranda:2011rt}, we make the assumption that the $1-2$ and $2-1$ entries in $M_{u,d}$ are equal leading to
\begin{equation}
\label{eq: Harayamaquarktheoretical}
 \hat{M}_{u,d}=m_{t,b}\left(\begin{array}{ccc}
                             0               & q_{u,d}/y_{u,d}      & 0  \\
                             q_{u,d}/y_{u,d} & 0                    & b_{u,d}(y_{u,d}) \\
                             0               & d_{u,d}(y_{u,d})     & y_{u,d}^{2}
                            \end{array}\right)
\end{equation}
where $y_{u,d}$ are free parameters and
\begin{eqnarray}
          p_{u,d} &=& \dfrac{m_{u,d}^{2}+m_{c,s}^{2}}{m_{t,b}^{2}} \ , \\
          q_{u,d} &=& \dfrac{m_{u,d}m_{c,s}}{m_{t,b}^{2}} \ , \\
 b_{u,d}(y_{u,d}) &=& \sqrt{\dfrac{p_{u,d}+1-y_{u,d}^{4}-R_{u,d}(y_{u,d})}{2}-\left( \dfrac{q_{u,d}}{y_{u,d}} \right)^{2}} \ , \\
 d_{u,d}(y_{u,d}) &=& \sqrt{\dfrac{p_{u,d}+1-y_{u,d}^{4}+R_{u,d}(y_{u,d})}{2}-\left( \dfrac{q_{u,d}}{y_{u,d}} \right)^{2}} \ ,
\end{eqnarray}
with
\begin{eqnarray}
 R_{u,d}(y_{u,d})=\bigl( (1+p_{u,d}-y_{u,d}^{4})^{2}-4(p_{u,d}+q_{u,d}^{4})+8q_{u,d}^{2}y_{u,d}^{2} \bigr)^{1/2} \ .
\end{eqnarray}
$\hat{M}_{u,d}$ are real matrices arising from the phase factorization of $M_{u,d}$
\begin{equation}
 M_{u,d}=P_{u,d}^{*}\hat{M}_{u,d}P_{u,d},
\end{equation}
with $P_{u,d}$ being diagonal phase matrices such that $P=P_{u}P_{d}^{*}=\text{diag}(1,e^{i\beta_{ud}},e^{i\alpha_{ud}})$, $\beta_{ud}=\beta_{u}-\beta_{d}$ and $\alpha_{ud}=\alpha_{u}-\alpha_{d}$. Therefore, four parameters $y_{u,d}$, $\beta_{ud}$ and $\alpha_{ud}$  have to be chosen to fit the CKM matrix,
\begin{equation}
 V_{CKM}=\mathcal{O}_{u}^{T}P\mathcal{O}_{d},
\end{equation}
where $\mathcal{O}_{u,d}$ diagonalize $\hat{M}_{u,d}\hat{M}_{u,d}^{T}$
\begin{equation}
 \mathcal{O}_{u,d}^{T}\hat{M}_{u,d}\hat{M}_{u,d}^{T}\mathcal{O}_{u,d}=\text{diag}(m_{u,d}^{2},m_{c,s}^{2},m_{t,b}^{2}).
\end{equation}
We show the numerical results that reproduce the experimental values in section~\ref{sec:numerical}.
\subsubsection{Charged Leptons}
\label{subsubsec:chargedlepton}

The charged lepton masses are similarly obtained from the Yukawa interactions on the IR brane
\begin{align}
\label{eq:yukawa-chargedlepton}
 -\mathcal{L}_{\text{Yukawa}}^{e} &= \int \text{d}y\dfrac{\sqrt{-g}~\delta(y-\pi R)}{\Lambda}\biggl[ (\gamma_{e}^{1})_{ij}\bar{L}_{Li}(x^{\mu},y)\Phi_{1}(x^{\mu})E_{Rj}(x^{\mu},y) \biggr. \notag \\
                                  &+ \biggl. (\gamma_{e}^{2})_{ij}\bar{L}_{Li}(x^{\mu},y)\Phi_{2}(x^{\mu})E_{Rj}(x^{\mu},y) \biggr].
\end{align}
After EWSB we obtain
\begin{equation}
\label{eq:Z4chargedleptonmassmatrix}
 M_{e}=m_{\tau}\left(\begin{array}{ccc}
                      0                                      & \Gamma_{12}^{e}\langle \Phi_{1}\rangle & 0 \\
                      \Gamma_{21}^{e}\langle \Phi_{1}\rangle & 0                                      & \Gamma_{23}^{e}\langle \Phi_{2}\rangle \\
                      0                                      & \Gamma_{32}^{e}\langle \Phi_{2}\rangle & \Gamma_{33}^{e}\langle \Phi_{1}\rangle
                     \end{array}\right).
\end{equation}
The analog of Eq.~(\ref{eq:4DYukawaquarks}) for charged leptons is
\begin{equation}
 \label{eq:4DYukawachargedleptons}
 \Gamma_{ij}^{e}=(\gamma_{e})_{i,j}\dfrac{k}{\Lambda}\sqrt{\dfrac{(1/2-c_{Li}^{L})(1/2-c_{Rj}^{e})}{[e^{2\pi kR(1/2-c_{Li}^{L})}-1][e^{2\pi kR(1/2-c_{Rj}^{e})}-1]}}e^{ \left(1-c_{Li}^{L}-c_{Rj}^{e}\right)\pi kR}.
\end{equation}
We can parametrize the $M_{e}$ matrix \footnote{We do not assume $(\hat{M}_{e})_{12}=(\hat{M}_{e})_{21}$ as we did in the quark sector.} following again~\cite{Harayama:1996am}, and so $M_{e}$ has dependence on both dimensionless parameters $y_{e}$ and $z_{e}$
\begin{equation}
\label{eq: Harayamachargedleptontheoretical}
 \hat{M}_{e}=m_{\tau}\left(\begin{array}{ccc}
                            0                  & q_{e}z_{e}/y_{e}  & 0 \\
                            q_{e}/(y_{e}z_{e}) & 0            & \sqrt{B_{e}(y_{e},z_{e})} \\
                            0                  & \sqrt{D_{e}(y_{e},z_{e})} & y_{e}^{2}
                           \end{array}\right) \ ,
\end{equation}
where $p_{e}$ and $q_{e}$ are the analogs of $p_{u,d}$ and $q_{u,d}$ for charged leptons and
\begin{eqnarray} \label{BD} \nonumber
 B_{e}(y_{e},z_{e}) &=& \dfrac{1}{2}\left\{ p_{e}+1-y_{e}^{4}\pm \sqrt{(1-p_{e}+y_{e}^{4})^{2}-4(q_{e}^{2}-y_{e}^{2}z_{e}^{2})(q_{e}^{2}-y_{e}^{2}/z_{e}^{2}}) \right\}-\dfrac{q_{e}^{2}}{y_{e}^{2}z_{e}^{2}} \\
 D_{e}(y_{e},z_{e}) &=& \dfrac{1}{2}\left\{ p_{e}+1-y_{e}^{4}\mp \sqrt{(1-p_{e}+y_{e}^{4})^{2}-4(q_{e}^{2}-y_{e}^{2}z_{e}^{2})(q_{e}^{2}-y_{e}^{2}/z_{e}^{2}}) \right\}-\dfrac{q_{e}^{2}z_{e}^{2}}{y_{e}^{2}} \ .
\end{eqnarray}

Observe we can have two different solutions: the plus (minus) case is obtained taking the plus (minus) and minus (plus) signs in $B_{e}$ and $D_{e}$ respectively. Depending on which solution we choose we have two different regions for the mathematically allowed values of  $z_{e}$ and $  y_{e}$, Region I (plus case) and II (minus case) respectively. Each region is constrained by the non -- negative real values of $B_{e}$ an $D_{e}$.

The matrix $\hat{M}_{e}$ allows us to obtain the lepton $c$~-- parameters by comparison to Eq.~(\ref{eq:Z4chargedleptonmassmatrix}). Under the assumption that the phases in the charged lepton sector are zero, there are only two parameters left, $y_{e}$ and $z_{e}$, whose actual values will be set by the neutrino sector results.

\subsubsection{Neutrino sector}
\label{subsubsec:neutrino}

Since there are no right-handed neutrinos present in the model, neutrino masses are generated radiatively~\footnote{Using the well known dimension five Weinberg operator leads to unacceptable large values for the neutrino masses. This could be remedied by allowing unnatural small values of the dimensionless $\gamma$ parameters, in contradiction to the philosophy of the general scenario. We therefore stick to the 4D renormalizable argument in~\cite{Aranda:2011rt} and consider the radiative mechanism.} as in~\cite{Aranda:2011rt}. In the present extradimensional setting this mechanism is going to set two of the lepton $c_{Li}^{L}$-parameters. As can be noted in the next section, their values turn out to be $\mathcal{O}(1)$, as any localization $c$~-- parameter must be.

The radiative mechanism introduces a cubic Lepton number violating scalar interaction among the two SU(2)$_L$ doublets and the charged scalar $h$, as well as the Zee operator coupling the left-handed lepton doublet to the singlet charged scalar~\cite{Zee:1980,Babu:1988qv}. The Randall-Sundrum geometry enhances this cubic scalar interaction as well as the Zee operator with factors proportional to the extradimensional profiles of the fermions and, if allowed to propagate through the bulk, those of the scalars. In the scenario where both scalar  Higgs doublets are confined to the IR brane and the scalar singlet $h$ is a bulk field, the cubic scalar operator is an interaction on the TeV brane which looks like
\begin{equation}
 -\mathcal{L}_{\Phi \Phi h}=\int \text{d}y\sqrt{-g}\delta(y-\pi R)\tilde{\lambda}_{\alpha \beta}\epsilon_{ij}\Phi_{i}^{\alpha}(x)\Phi_{j}^{\beta}(x)h(x,y) \ ,
\end{equation}
where  $i$, $j$ are SU(2) indices and $\tilde{\lambda}_{\alpha \beta}$ is antisymmetric and with mass dimension $+1/2$. Under the ZMA approach this term acquires the form
\begin{equation}
  -\mathcal{L}_{\Phi \Phi h}=\int_{-\pi R}^{+\pi R}\text{d}y~e^{-4k|y|}\delta(y-\pi R)\lambda_{\alpha \beta}\sqrt{\Lambda}\epsilon_{ij}\Phi_{i}^{\alpha}(x)\Phi_{j}^{\beta}(x)\dfrac{h^{(0)}(x)f_{h}^{(0)}(y)}{\sqrt{2\pi R}}\bigl( e^{\pi kR} \bigr)^{2} \ ,
\end{equation}
with the dimensionless antisymmetric coupling $\lambda_{\alpha \beta}$ of $\mathcal{O}(1)$ (note also the inclusion of a canonically-normalizing factor for each one of the Higgs doublets). Thus, for an UV-peaked bulk $h$ field (whose approximated profile and normalization factor are  shown 
in appendix~\ref{appendix:profiles}) the effective cubic operator acquires the form
\begin{equation}
 -\mathcal{L}_{\Phi \Phi h}=\lambda_{\alpha \beta}\epsilon_{ij}\Phi_{i}^{\alpha}(x)\Phi_{j}^{\beta}(x)\cdot \eta_{\lambda}^{(UV)} \ ,
\end{equation}
where
\begin{equation}
 \eta_{\lambda}^{(UV)}=\sqrt{\Lambda}\sqrt{k|b_{-}-1|}e^{(b_{-}-2)\pi kR}
\end{equation}
has the correct mass dimension: $[\eta_{\lambda}^{(UV)}]=+1$ and $b_{-}<1$. An UV-peaked bulk $h$ (again in the ZMA) also modifies the Zee operator
\begin{equation}
 \mathcal{L}_{LLh}=\int \text{d}y\sqrt{-g}\tilde{\kappa}^{ab}\epsilon_{ij}\overline{(L_{Li}^{a})^{c}}(x,y)L_{Lj}^{b}(x,y)h^{*}(x,y) \ ,
\end{equation}
(where $[\tilde{\kappa}^{ab}]=-1/2$ and antisymmetric by the Pauli principle) and leaves it as
\begin{equation}
 \mathcal{L}_{LLh}=\int_{-\pi R}^{+\pi R}\text{d}y~e^{-4k|y|}\dfrac{\kappa^{ab}}{\sqrt{\Lambda}}\epsilon_{ij}\overline{(L_{Li}^{a(0)})^{c}}(x)L_{Lj}^{b(0)}(x)h^{(0)*}(x)\dfrac{f_{Li}^{a(0)}(y)}{\sqrt{2\pi R}}\dfrac{f_{Lj}^{b(0)}(y)}{\sqrt{2\pi R}}\dfrac{f_{h}^{(0)*}(y)}{\sqrt{2\pi R}} \ ,
\end{equation}
where $\kappa^{ab}$ is now dimensionless (and antisymmetric). The effective Zee operator then looks like
\begin{equation}
 -\mathcal{L}_{LLh}=\kappa^{ab}\epsilon_{ij}\overline{(L_{Li}^{a(0)})^{c}}(x)L_{Lj}^{b(0)}(x)h^{*(0)}(x)\cdot \eta_{\kappa}^{(UV)} \ ,
\end{equation}
where $\eta_{\kappa}^{(UV)}$ is dimensionless and given by 
\begin{equation}
 \eta_{\kappa}^{(UV)}=\dfrac{k^{3/2}}{\sqrt{\Lambda}}\sqrt{\dfrac{(\tfrac{1}{2}-c_{La}^{L})(\tfrac{1}{2}-c_{Lb}^{L})|b_{-}-1|}{[e^{2\pi kR(1/2-c_{La}^{L})}-1][e^{2\pi kR(1/2-c_{Lb}^{L})}-1]}}\int_{-\pi R}^{+\pi R}\text{d}y~e^{(b_{-}-c_{La}^{L}-c_{Lb}^{L})k|y|} \ .
\end{equation}
Following the expressions in~\cite{Aranda:2011rt} for the 4D Majorana neutrino mass matrix and replacing the effective Yukawa and scalar couplings, we find that the Majorana neutrino mass matrix entries are
\begin{eqnarray} \label{eq:mll-entries} 
m_{\nu_{e}\nu_{e}} &=& 2a_{e}\kappa^{31}\eta_{\kappa}^{(UV)}m_{\tau \mu}\lambda_{12}\eta_{\lambda}^{(UV)}\langle \Phi_{2}\rangle F(m_{\Phi}^{2},m_{h}^{2}), \\ 
m_{\nu_{\mu}\nu_{\mu}} &=& 0, \\ 
m_{\nu_{\tau}\nu_{\tau}} &=& 2{b}_{e}'\kappa^{13}\eta_{\kappa}^{(UV)}m_{e\mu}\lambda_{21}\eta_{\lambda}^{(UV)}\langle \Phi_{1}\rangle F(m_{\Phi}^{2},m_{h}^{2}), \\ 
m_{\nu_{e}\nu_{\mu}}&=&m_{\nu_{\mu}\nu_{e}} = 2b_{e}\kappa^{31}\eta_{\kappa}^{(UV)}m_{\tau \tau}[\lambda_{21}\eta_{\lambda}^{(UV)}\langle \Phi_{1}\rangle F(m_{\Phi}^{2},m_{h}^{2}), \\ 
m_{\nu_{e}\nu_{\tau}}&=&m_{\nu_{\tau}\nu_{e}} = 2\biggl( a_{e}\kappa^{13}\eta_{\kappa}^{(UV)}m_{e\mu}\lambda_{12}\eta_{\lambda}^{(UV)}\langle \Phi_{2}\rangle+b_{e}'\kappa^{31}\eta_{\kappa}m_{\tau \mu}\lambda_{21}\eta_{\lambda}^{(UV)}\langle \Phi_{1}\rangle \\ 
                                                 &+& c_{e}\kappa^{31}\eta_{\kappa}^{(UV)}m_{\tau \tau}\lambda_{12}\eta_{\lambda}^{(UV)}\langle \Phi_{2}\rangle \biggr)F(m_{\Phi}^{2},m_{h}^{2}), \\ 
m_{\nu_{\mu}\nu_{\tau}}&=&m_{\nu_{\tau}\nu_{\mu}} = 0,
\end{eqnarray}
where $F(m_{\Phi}^{2},m_{h}^{2})$ (accounting for the scalar loop factor) is a function depending on the scalar masses of the charged Higgs, $m_{\Phi}$, and of the singlet scalar $m_{h}$ through
\begin{equation}
 F(m_{\Phi}^{2},m_{h}^{2})=\dfrac{1}{16\pi^{2}}\dfrac{1}{m_{\Phi}^{2}-m_{h}^{2}}\log{\dfrac{m_{\Phi}^{2}}{m_{h}^{2}}}.
\end{equation}
The parameters $a_{e}$, $b_{e}$, $b_{e}'$, and $c_{e}$ belong to the Yukawa matrix for charged leptons
\begin{equation*}
 Y_{e}=\left(\begin{array}{ccc}
                            0 & \Gamma_{12}^{e} & 0 \\
              \Gamma_{21}^{e} & 0               & \Gamma_{23}^{e} \\
                            0 & \Gamma_{32}^{e} & \Gamma_{33}^e
             \end{array}\right)\equiv \left(\begin{array}{ccc}
                                                  0 & a_{e}  & 0 \\
                                             a_{e}' & 0      & b_{e} \\
                              0 & b_{e}' & c_{e}
                           \end{array}\right)
\end{equation*}
and the entries $m_{ij}$, $i,j=e,\mu,\tau$ are those of the  matrix $M_{e}$.

Now, since $\kappa_{ij}$ and $\lambda_{ij}$ are antisymmetric we can rewrite the neutrino mass matrix entries as 
\begin{eqnarray} 
m_{\nu_e\nu_e} &=& -(m_{e\mu} m_{\tau\mu}\tan\beta)  \mathcal{C}, \\  
m_{\nu_{\mu}\nu_{\mu}} &=& 0, \\  
m_{\nu_{\tau}\nu_{\tau}} &=& -\frac{m_{\tau\mu} m_{e\mu}}{\tan\beta}\mathcal{C}, \\  
m_{\nu_e\nu_{\mu}} &=& m_{\nu_{\mu}\nu_{e}}=\frac{m_{\mu\tau} m_{\tau\tau}}{{\tan\beta}}\mathcal{C}, \\  
m_{\nu_e\nu_{\tau}} &=& m_{\nu_{\tau}\nu_e}=\left( m_{e\mu}^2\tan\beta+\frac{{m_{\tau\mu}}^2}{{\tan\beta}}-{m_{\tau\tau}^2} {\tan\beta}\right)\mathcal{C}, \\  
m_{\nu_{\mu}\nu_{\tau}} &=& m_{\nu_{\tau}\nu_{\mu}}=0 ,
\end{eqnarray}
where $\mathcal{C}$  is a common factor with  dimension of inverse of mass
\begin{equation}\label{common}
 \mathcal{C}=2  \kappa^{13}\eta_{\kappa}^{UV} \lambda_{12} \eta_{\lambda}^{UV} F(m_{\Phi}^2,m_h^2),
\end{equation}
and 
\begin{equation}
 \tan\beta=\frac{\langle \Phi_2 \rangle}{\langle \Phi_1 \rangle}.
\end{equation}

\section{Numerical Analysis}
\label{sec:numerical}

\subsection{Lepton sector}

To perform the numerical analysis in the lepton sector we used the experimental data at $3\sigma$ from the global neutrino data analysis in~\cite{Schwetz:2011zk} 
\begin{equation}
 \begin{array}{|c|c|c|}
\hline
& \text{Best Fit Value}& 3\sigma \text{ range}\\
\hline
\sin^2 \theta_{12} & 0.312 & 0.27-0.36\\
\sin^2\theta_{23}& 0.52 & 0.39-0.64\\
\Delta m^2_{21} [10^{-5} \text{eV}^2]&7.59 & 7.09-8.19\\
 \Delta m^2_{32} [10^{-3} \text{eV}^2 ]& 2.50 & 2.14-2.76\\
               &-(2.40)&-(2.13-2.67) \\                 
\hline
 \end{array}
\end{equation}
with  $\delta_{CP}=0$ and normal (inverted) hierarchy
and the recently Daya Bay results (confirmed at $5\sigma$)~\cite{An:2012eh}
\begin{equation}
 \sin^2 2\theta_{13}=0.092\pm 0.017,
\end{equation}
which can be rewritten as 
\begin{equation}
 \sin^2 \theta_{13}=0.0235 \pm 0.0045.
\end{equation}
By convenience in the analysis we also define the following range for the  neutrino squared mass differences ratio (at $3\sigma$)
\begin{equation}\label{ratio}
\text{NH (IH):  } 0.027(0.028) < \left |\frac{\Delta m^2_{21}}{\Delta m^2_{32}}\right| < 0.035(0.036),
\end{equation}
obtained by summing in quadrature the relative errors of $\Delta m^2_{21}$ and $|\Delta m^2_{32}|$. The constraint over the sum of neutrino masses, $\sum m_{\nu_i}<0.29$ eV presented recently in~\cite{RiemerSorensen:2011fe} is also considered.

The charged leptons masses used in the analysis are those given by the central values in~\cite{PDG}
\begin{eqnarray}
 m_e&=&0.511\times 10^{-3} \text{ GeV},\\
m_{\mu}&=&0.1056  \text{ GeV},\\
m_{\tau}&=&1.776  \text{ GeV}.
\end{eqnarray}

As in reference~\cite{Kadosh:2010rm} we take the 5D scale to be $\Lambda=k=M_{Planck}$ where $M_{Planck}=2.44\times10^{15}$~TeV is the reduced Planck mass, and the effective scale $k e^{-\pi k R}\approx 6.89$~TeV generated by $\pi k R=33.5$. For the other parameters involved in the neutrino mass sector we take $m_{\Phi}= 0.5$~TeV (recall this is the mass of the {\it charged} Higgses in the loop), $m_h = M_{Planck}$, $\kappa^{13}= 1=-\kappa^{31}$, and $\lambda^{12}= 1=-\lambda^{21}$.  Note that since $h$ is UV-peaked its mass is taken to its natural value of $M_{Planck}$. We note however that the model can reproduce the neutrino sector even in the case of a very light $h$-field (or even an IR-peaked $h$ field) due to the fact that $m_h$ only enters through the common factor $\mathcal{C}$ in Eq.~\eqref{common}, and thus does not affect the diagonalization. Its only possible effect is in the absolute size of the neutrino mass matrix entries and it corresponds to acceptable changes of $O(1)$.

From the expressions for the neutrino mass matrix entries we observe that the only  $c_{Li}^L$ coefficients involved are $c_{L1}^L$ and $c_{L3}^L$, and are taken to be $c_{L1}^L=0.8$ and  $c_{L3}^L=0.55$ (this is only a choice and corresponds to similar values used in~\cite{Kadosh:2010rm}). We also observe that all these parameters are contained in the $\mathcal{C}$ factor, together with the parameter  $b_{_-} = 2- \sqrt{4+a_{_-}}$, and in consequence they do not affect the diagonalization matrix. Thus, the lepton mixing matrix $U_{PMNS}$ defined by $U_{PMNS}=U_{L}^{\dag}U_{\nu}$, where $U_{L}$ and $U_{\nu}$ are the diagonalization matrices in the charged lepton and neutrino sectors respectively, only depends on $\tan \beta$ and the charged leptons mass matrix entries.

We use the standard parametrization of the  $U_{PMNS}$ given in~\cite{PDG}
\begin{equation}
 V=U_{PMNS} P,
\end{equation}
where
\begin{equation}
\label{Umns}
U_{PMNS}= \left(\begin{array}{ccc}
c_{13}c_{12}&c_{13}s_{12}&s_{13}e^{-\imath\delta_{CP}}\\
-c_{23}s_{12}-s_{23}s_{13}c_{12}e^{\imath\delta_{CP}}&c_{23}c_{12}-s_{23}s_{13}s_{12}e^{\imath\delta_{CP}} &s_{23}c_{13}\\
s_{23}s_{12}-c_{23}s_{13}c_{12}e^{\imath\delta_{CP}} &-s_{23}c_{12}-c_{23}s_{13}s_{12}e^{\imath\delta_{CP}} &c_{23}c_{13}\\
\end{array}\right),
\end{equation}
with $c_{ij}\equiv\cos \theta_{ij}$, $s_{ij}\equiv\sin\theta_{ij}$ and where $\delta_{CP}$ is the Dirac-CP violating phase in the neutrino sector which we assume is zero. $P=\text{diag}(1,e^{\imath \alpha_{21}},e^{\imath \alpha_{31}})$ is a diagonal phase matrix containing the two Majorana CP violating phases.

The first step in our analysis consists of a scan over the mathematically allowed regions for the $z_e$ and $y_e$ parameters of the charged lepton mass matrix $M_e$ that lead to positive values for $B_e$ and $D_e$ in Eq.~(\ref{BD}). The next step is to scan over values for $\tan\beta$ values that lead to the allowed neutrino mass differences ratio and mixing angles. 
Finally, once the ranges for $z_e$, $y_e$ and  $\tan\beta$ values that give the right angles and ratios are determined, we fit the $a_{_-}$ values (or $b_{_-}$) required to obtain the neutrino masses satisfying the constraint over the sum of neutrino masses~\cite{RiemerSorensen:2011fe}.

Performing the scan over $\tan\beta$ values from $0.1$ to $1.0$ (in steps of $0.1$) and from $1$ to $20$ (in steps of $1$)  we find that in Region I (plus case) the only value that works is $\tan\beta=2.0$ while in Region II (minus case) $\tan\beta=0.3$ and  $\tan\beta=0.6$ give acceptable values. 

\begin{figure}[h!]
\includegraphics[width=0.55\textwidth]{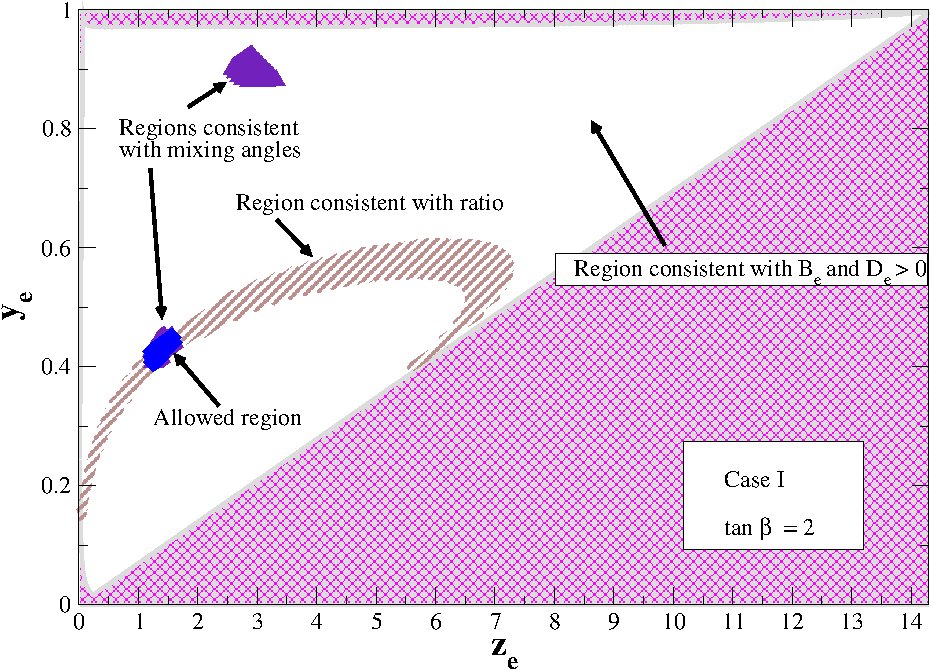}
\includegraphics[width=0.55\textwidth]{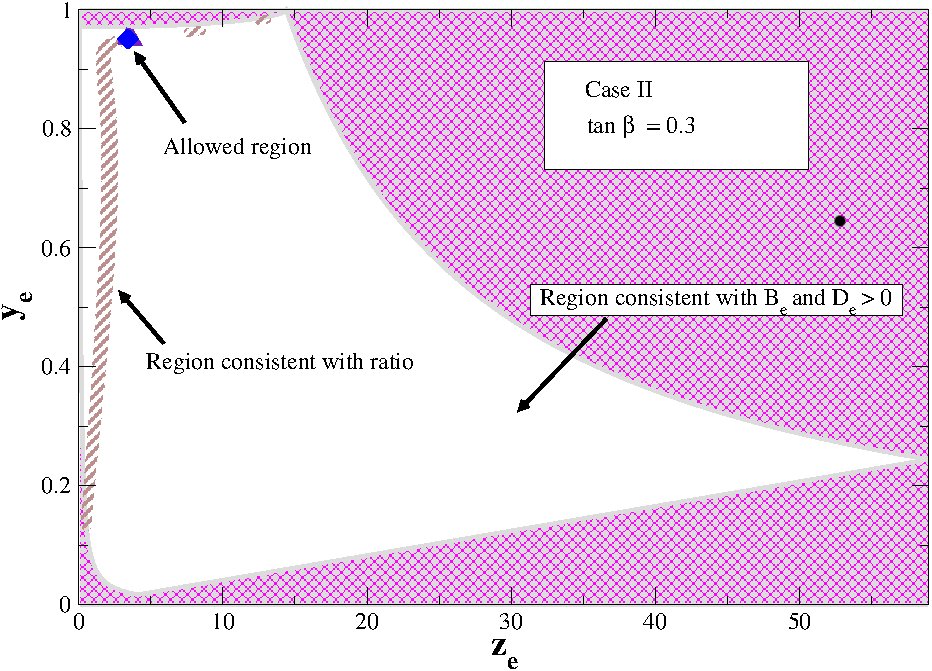}
\includegraphics[width=0.55\textwidth]{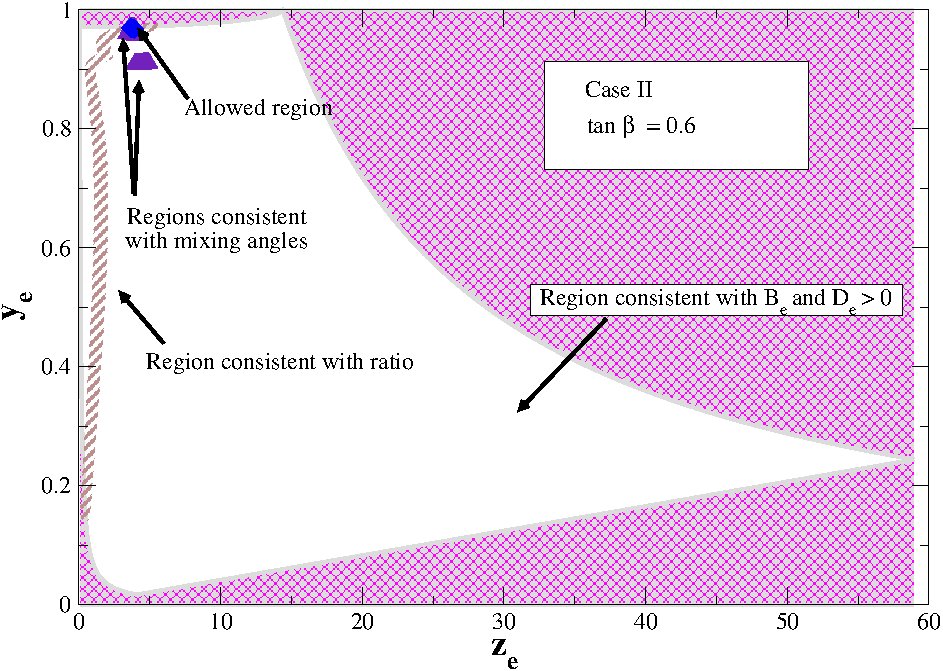}
\caption{Consistent regions of parameter space in the $z_e - y_e$ plane for Case I and $\tan\beta = 2$ (upper plot), Case II and $\tan\beta = 0.3$ (middle), and Case II and $\tan\beta = 0.6$ (bottom). We show the complete region consistent with positive values for $D_e$ and $B_e$, the region consistent with the neutrino mass squared differences ratio, the regions consistent with neutrino mixing angles, and the {\it allowed region} (see text). Note that for Case II and $\tan\beta = 0.3$ the region consistent with mixing angles is basically the same as the allowed region.}
\label{fig:fig1}
\end{figure}

Figure~\ref{fig:fig1} shows the parameter space obtained in this process. Note that in all cases there is a large region in the $z_e - y_e$ plane consistent with positive values for $B_e$ and $D_e$. However the allowed region, consistent with all experimental information, is considerably reduced to small ranges in this parameter space. In the plots we label by {\it allowed region} the region of parameter space consistent with all experimental data, including the constraint on the sum of neutrino masses~\cite{RiemerSorensen:2011fe}.

We note that, as found in~\cite{Aranda:2011rt}, only inverted hierarchy for the neutrino masses is obtained in this model and that the resulting Majorana phases are found to be $\alpha_{21}=\alpha_{31}=\pi$.

It is important to note that throughout the analysis all the $\gamma_{ij}^{e}$ have been taken to be of $\mathcal{O}(1)$. In fact, except for $\gamma_{12}^{e}$ and $\gamma_{32}^{e}$ which are used in the fit and take values of $\mathcal{O}(1)$ in all cases, all other are set equal to $1$. The same is true for all the profile $c$~-- coefficients: all are of $\mathcal{O}(1)$. 

\subsection{Quark sector}
In~\cite{Aranda:2011dx} the values for the parameters $y_{u,d}$, $\beta_{ud}$ and $\alpha_{ud}$ leading to a correct CKM matrix $|V_{CKM}|$ were found to be $y_{u}=0.9964$, $y_{d}=0.9623$, $\beta_{ud}=1.4675$ and $\alpha_{ud}=1.9560$. The corresponding mass matrices (in TeV) read
\begin{eqnarray}
 \label{harayamaquarknum}
 \hat{M}_{u} &=& \left(\begin{array}{ccc}
                       0            & 0.0000552118 & 0 \\
                       0.0000552118 & 0            & 0.0118527 \\
                       0            & 0.0183451    & 0.17152
                      \end{array}\right) \\
 \hat{M}_{d} &=& \left(\begin{array}{ccc}
                       0            & 0.0000235694  & 0 \\
                       0.0000235694 & 0             & 0.000268666 \\
                       0            & 0.00158321    & 0.0039362
                      \end{array}\right).
\end{eqnarray}
Comparing these matrices with Eq.~(\ref{eq:Z4quarkmassmatrix}) and using the values for $\tan\beta$ found to work in the neutrino sector, we performed a fit to the $c$~-- profile parameters and $\gamma$'s. We find that it is possible to obtain solutions with all parameters of $\mathcal{O}(1)$. In particular, for $\tan\beta=0.3$, experimental agreement is found for at least the following set of values $(c_{L1}^{Q},c_{L2}^{Q},c_{L3}^{Q}) = (0.7, 0.491253, 0.128276)$, $(c_{R1}^{u},c_{R2}^{u},c_{R3}^{u}) = (0.666249, 0.342679, 0.737729)$, $(c_{R1}^{d},c_{R2}^{d},c_{R3}^{d}) = (0.690378, 0.468589, 0.591601)$, $(\gamma^u_{ij}) = 1_{3\times 3}$, $\gamma_{32}^{d} = 0.101081$, $\gamma_{33}^{d} = 2.02488$, and all other $\gamma$'s equal to $1$.

Thus, using dimensionless parameters of order one and dimensionful parameters associated to the scale of the setup, we find that it is possible to reproduce all observed masses and mixing angles in both the quark and lepton sector. Furthermore we accomplish this with a minimal set of additions and only left-handed neutrinos. It is important to recall that all results presented in this work have been obtained under the ZMA scheme and thus neglecting (although safely~\citep{Kadosh:2010rm}) possible contributions from KK modes. Yet another possible source of contributions that must be investigated corresponds to contributions from higher order operators consistent with the gauge and flavor symmetries. In our case, the smallest higher order operators that could contribute are of the general form $\Phi \Phi \Phi \bar{Q}_LQ_R$ (and similarly for leptons). We find that the gauge and flavor invariant contributions, denoted by $\delta m_{ij}$, to the zero entries in the mass matrices satisfy $\delta m_{ij}/m_{lk} \sim 10^{-7} - 10^{-11}$, where the $m_{lk}$ denote the values of the non-zero entries. For the case $ij=lk$, corresponding to contributions to the non-zero entries, we find $\delta m_{ij}/m_{ij} \sim 10^{-4}$. Thus, these contributions can be safely ignored. 

Before concluding we make some remarks to be taken in full consideration in a future work. Our model has a potentially interesting scalar phenomenology that requires a complete study, including its vacuum stability and possible collider signals. Another important and related phenomenological issue is the presence of flavor changing neutral current (FCNC) effects. In our model there are two different possible contributions to FCNC that must be taken into consideration. The first one comes from the fact that we are dealing with a two (flavored) Higgs doublet model in which each Higgs couples to both the up and down quark sector. This can result in tree-level FCNC~\cite{Atwood:1996vj,Azatov:2009na} that need to be analyzed and will induce constraints on the parameter space in the scalar potential. The second possibility is associated to the presence of mixing between the degrees of freedom of the effective 4D theory and their KK excitations. This is a problem that all extra dimensions models have to face. One has to guarantee that the corrections induced by this mixing will keep the agreement with observations forcing a lower bound, of order a few TeV, on the KK scale. In our model, the use of a custodial symmetry allows to consider a lower bound on the first KK mass to that imposed by electroweak precision measurements~\cite{Agashe:2003zs} (see also~\cite{Chen:2008qg,Bauer:2011ah,Casagrande:2010si}).
Furthermore if the flavor pattern of the Yukawa couplings holds in the 5D theory due to a flavor symmetry, an increased alignment between the 4D fermion mass matrix and the Yukawa and gauge couplings is obtained suppressing the amount of flavor violation induced by the interactions with KK states. Moreover, as pointed out in~\cite{Kadosh:2010rm}, a bulk flavor symmetry might also induce a cancellation of observable phases,  therefore the dominant new physics contributions to the neutron and electron dipole moments and/or to $\epsilon'/\epsilon_K$ will vanish.

\section{Conclusions}
\label{sec:conclusions}
A five-dimensional warped extension of the Standard Model is considered, where the space-time background is taken to be the Randall-Sundrum model 1, i.e. a warped extra-dimension of Planck size in an Anti-de Sitter bulk.  We consider a Zero-Mode Approximation for the four-dimensional effective action one obtains as a dimensional reduction of the higher-dimensional theory and that, in the present case, constitutes an extension of the SM. In the model we study above almost all the fields in the 4D effective action are zero modes of bulk fields, in particular each left- respectively right-handed fermion mode has its five-dimensional counterpart; only the Higgs fields are purely four-dimensional and live on the IR brane (i.e. on the 4D plane located at the orbifold fixed point $y=\pi R$) and, unlike the bulk fields, have no KK excitations. Bulk fermion fields come about with a (Planck size) mass term parametrized by a real number $c$ which determines the shape of the extra-dimensional profile of the field. The Yukawa interaction between Higgses and fermions is given in terms of overlapping integrals of zero-mode profiles for the fields involved and thus depends upon the aforementioned $c$ real parameters along with numerical coupling constant (hermitian) matrices $\gamma$'s. In order to obtain a NNI-type quark mass matrix, we advocate the presence of a discrete symmetry upon which all fields are charged: the cyclic group $Z_4$ is the smallest Abelian group consistent with the aforementioned texture. Also for the charged lepton sector the discrete symmetry fixes the form of mass matrix in a similar way as for the quark sector. For the quark sector (and similarly for the charged lepton sector) the numerical analysis goes as follows: first we consider the mass matrix parametrized {\em \`a la} Harayama in order for it to display NNI form.  The diagonalization of the real squared counterpart of the previous mass matrix leaves four parameters that can be fixed using the experimental values for the CKM. In turn, this allows to fix the entries of the mass matrix itself and, by comparison of such entries with the ones foreseen by the overlaps of extra-dimensional profiles, one can obtain the compactification parameters $c$'s and the Yukawa dimensionless entries $\gamma$´'s. All these parameters turn out to be of ${\mathcal O}(1)$.  For the charged leptons the parametrization goes slightly different and the (real) mass matrix depends upon two free parameters only constrained by the reality of the mass matrix: allowed parameter regions satisfying such constraints, as well as all experimental data, are graphically displayed. In the neutrino sector we choose not to introduce right-handed neutrinos in the model and generate neutrino masses radiatively. This is accomplished with the help of a bulk charged scalar field whose zero-mode is peaked on the UV brane (located at the fixed point $y=0$): such field also mediates the lepton number violation. With scalar field masses chosen to be $m_\Phi = 0.5$~TeV for the charged Higgses in the loops, and $m_h = M_{Planck}$~TeV, experimental data taken from $3\sigma$ global neutrino data analysis and Daya Bay results are matched by setting all $c$ parameters and $\gamma$ parameters to values of ${\mathcal O}(1)$. Although a complete phenomenological study of the model is under preparation, some comments regarding the scalar phenomenology of the model as well as its possible contributions to FCNC were also briefly discussed.
\acknowledgments{OC is grateful to the University of Colima and the INFN Bologna for hospitality and support while parts of this work were completed. CA, AA, and ADR thank the Universidad Aut\'onoma de Chiapas  for its hospitality while part of this work was done. The work of OC was partly funded by SEP-PROMEP/103.5/11/6653. AA acknowledges support from CONACYT under the program Estancias de Consolidaci\'on, grant No. 145378. }

\newpage
\begin{appendix}

\section{On the zero mode profiles}
\label{appendix:profiles}

We briefly review the computation of zero-mode profiles for some bulk
fields, on the RS1 background considered above; we basically follow
what is done in~\cite{Pomarol00,Huber:2000ie}. A bulk field satisfies a
second order differential equation of the form  
\begin{eqnarray}
\biggl[ e^{2\sigma} \eta^{\mu\nu} \partial_\mu \partial_\nu +e^{s
  \sigma} \partial_y \left(e^{-s \sigma} \partial_y\right) -M_{\Phi}^2
\biggr] \Phi(x,y)=0  \ ,
\end{eqnarray}
where $\Phi=\{A_\mu, \phi, e^{-2\sigma} \Psi_{L,R}\}$, $s=\{2,4,1\}$
and $M_{\Phi}^2=\{0,a k^2 +b\sigma'', c(c+1)k^2 -c \sigma''\}$;
note that compared to~\cite{Pomarol00,Huber:2000ie}'s we redefined $c \to
-c$ for the right-handed mode. Above $\sigma(|y|) = k|y|~ ({\rm modulo}\
2\pi R)$ and thus $\sigma' = k\epsilon(y)$ and $\sigma'' = 2k[
\delta(y) -\delta(y-\pi R)]$. Upon Kaluza-Klein decomposition one
gets
\begin{eqnarray}
  \Phi(x,y) =\frac1{\sqrt{2\pi R}} \sum_{n=0}^\infty \Phi^{(n)}(x) f_n(y) \ ,
\end{eqnarray}
with $\eta^{\mu\nu} \partial_\mu \partial_\nu \Phi^{(n)}(x) =m_n^2
\Phi^{(n)}(x),$ and orthonormality conditions 
\begin{eqnarray}
  \frac1{2\pi R} \int_{-\pi R}^{\pi R} dy~e^{2l_s\sigma} f_n(y)
  f_{n'}(y) = \delta_{n,n'}~,\quad l_s= (0,-1,-3/2) 
\end{eqnarray}
so that
\begin{eqnarray}
\biggl[ -e^{s
  \sigma} \partial_y \left(e^{-s \sigma} \partial_y\right) +M_{\Phi}^2
\biggr] f_n(y)=m_n^2 f_n(y)~.  
\end{eqnarray}
Here we only concentrate on the zero modes for which $m_0 =0$. 
\subsection*{Vector field}
For the vector field it is immediate to realize that the only possible
zero-mode profile is constant and since the measure is also trivial in this
case, we simply have $f_0(y) =1$.
\subsection*{Scalar field}
For a bulk scalar field we have
\begin{eqnarray}
\biggl[ -e^{4
  \sigma} \partial_y \left(e^{-4 \sigma} \partial_y\right) +ak^2 +b \sigma''
\biggr] f_0(y)=0  
\end{eqnarray}
that admits solutions
\begin{eqnarray}
  f_{0,\pm}(y) = \frac1{N_0} e^{b_{_\pm}\sigma}\,, \quad b_{_\pm} = 2\pm \sqrt{4+a} \ ,
\end{eqnarray}
and the normalization factor is given by 
\begin{eqnarray}
  \label{eq:N0}
  N_{0,\pm} = \left[ \frac{e^{2(b_{_\pm}-1)\pi kR}-1}{2 \pi k
    R(b_{_\pm}-1)}\right]^{1/2} \approx  \frac1{\sqrt{2\pi k
    R |b-1|}}\left\{ \begin{array}{ll} 
e^{(b-1)\pi kR}\,,\quad\ & b=b_+\\[1mm]
1\,, \quad\ & b=b_-
\end{array} \right.
\end{eqnarray}
where in the last expression we took into account that $k \pi R >>1$
and $b_{_+} >1$, $b_{_-}<1$. Hence, including the measure factor the
zero-mode profiles read
\begin{eqnarray}
  \label{eq:scalar-profile}
  \frac1{\sqrt{2\pi R}} e^{-\sigma} f_0(y) \approx \sqrt{|b - 1| k}
\left\{ \begin{array}{ll} 
e^{(b_{_+}-1)k(|y|-\pi R)}\,,\quad\ & b=b_+\\[1mm]
e^{(b_{_-}-1)k|y|}\,, \quad\ & b=b_-
\end{array} \right. 
\end{eqnarray}
so that for $b=b_{_+}$ the profile is peaked about the IR brane
($y= \pi R$) and for $b=b_{_-}$ the profile is peaked about the UV
brane ($y=0$).  
\subsection*{Fermionic field}
For a bulk fermionic field we follow~\cite{Huber:2000ie} with the
aforementioned renaming of parameter $c$ for the right-handed
mode. Hence  
\begin{eqnarray}
\biggl[ -e^{
  \sigma} \partial_y \left(e^{-\sigma} \partial_y\right) + c(c+1)k^2 -c \sigma''
\biggr] e^{-2\sigma}f_0(y)=0  
\end{eqnarray}
and 
\begin{eqnarray}
  \label{eq:fermion-profile}
  f_{0\,L,R}(y) =\frac1{N_0} e^{(2-c)\sigma} \ ,
\end{eqnarray}
with 
\begin{eqnarray}
N_0 = \left[\frac{e^{2\pi kR(1/2-c)} -1}{2\pi kR(1/2 -c)}\right]^{1/2} \approx \frac1{\sqrt{2\pi k
    R |1/2-c|}}\left\{ \begin{array}{ll} 
e^{(1/2-c)\pi kR}\,,\quad\ & c < 1/2\\[1mm]
1\,, \quad\ & c>1 /2
\end{array} \right.
\end{eqnarray}
Hence, including the measure factor the
zero-mode profiles read
\begin{eqnarray}
  \label{eq:fermionic-profile}
  \frac1{\sqrt{2\pi R}} e^{-3\sigma/2} f_0(y) \approx \sqrt{|1/2 - c| k}
\left\{ \begin{array}{ll} 
e^{(1/2-c)k(|y|-\pi R)}\,,\quad\ & c<1/2\\[1mm]
e^{(1/2-c)k|y|}\,, \quad\ & c>1/2
\end{array} \right. 
\label{eq:zero-modeF}
\end{eqnarray}
so that for $c<1/2$ the profile is peaked about the IR brane
($y= \pi R$) whereas for $c>1/2$ the profile is peaked about the UV
brane ($y=0$).  
\end{appendix}

\end{document}